\documentclass[11pt,a4paper]{article}

\usepackage{jcappub}

\usepackage{caption}
\usepackage{subcaption}
\usepackage{amssymb}
\usepackage{amsfonts}
\usepackage{amsmath}
\usepackage{graphicx}
\usepackage{dcolumn}
\usepackage{bm}

\title{Reference Frames and Black Hole Thermodynamics}
\author[a]{Franco Fiorini,}
\author[b]{P. A. Gonz\'{a}lez}
\author[c]{and Yerko V\'asquez}
\affiliation[a]{Departamento de Ingenier\'{i}a en Telecomunicaciones (CONICET) and Instituto Balseiro (UNCUYO), Centro At\'{o}mico Bariloche, Av. Ezequiel Bustillo 9500, CP8400, S. C. de Bariloche, R\'{i}o Negro, Argentina.}
\affiliation[b]{Facultad de
Ingenier\'{i}a y Ciencias, Universidad Diego Portales, Avenida Ej\'{e}rcito
Libertador 441, Casilla 298-V, Santiago, Chile.}
\affiliation[c]{Departamento de F\'isica, Facultad de Ciencias, Universidad de La Serena,\\
Avenida Cisternas 1200, La Serena, Chile.}
\emailAdd{francof@cab.cnea.gov.ar}
\emailAdd{pablo.gonzalez@udp.cl}
\emailAdd{yvasquez@userena.cl}
\abstract{In the context of the absolute parallelism formulation of General Relativity, and because of the fact that the scalar curvature can be written in purely torsional terms, it was known for a long time that a surface term based solely on the torsion tensor appears in the action. It was subsequently suggested that this term might play the role of the Gibbons-Hawking-York boundary term which, in turn, is associated to the free energy in the path integral approach, and then, to the black hole entropy by standard thermodynamic arguments. We show that the identification of the two boundary terms is rather incomplete, and that it strongly depends on the choice of the tetrad (frame) field used to reproduce a given metric. By considering variations of the tetrad field not necessarily subjected to Dirichlet-like conditions on the boundary surface, we find a class of frames adapted to the Schwarzschild spacetime in which the Gibbons-Hawking-York/torsion link is actually established, and conducing to the right black hole entropy without the need of any background subtraction. Remarkably, these frames are also responsible for the correct value of the gravitational energy as computed from the teleparallel energy-momentum pseudo-current.}
\arxivnumber{...}

\begin{document}
\maketitle

\section{Introduction}

It is well known that when General Relativity (GR) is formulated in a manifold $\mathcal{M}$ with boundary $\partial \mathcal{M}$, it requires the presence of a boundary counter-term in order to cancel the one coming from the Einstein-Hilbert (EH) action 
\begin{equation}
  \label{EHac}
    S_{EH} =\frac{1}{16 \pi}\int_{\mathcal{M}} d^4x \sqrt{g}\,R\,,
\end{equation}
where $g=-det(g_{\mu\nu})$ and $R$ is the scalar curvature (we are using units in which $G=c=\hbar=1$). For this reason, the Gibbons-Hawking-York (GHY) boundary term \cite{Gibbons:1976ue, York:1972sj}
\begin{equation}
  \label{terGH}
    S_{GHY} =\frac{\epsilon}{8 \pi }\oint_{\partial \mathcal{M}} d^3x \sqrt{h}\,K\,,
\end{equation}
is used as a counter-term to correct the EH action once variations of the metric $g_{\mu\nu}$ subjected to Dirichlet boundary conditions $\delta g_{\mu\nu}\mid_{\partial \mathcal{M}}=0$ on the boundary $\partial \mathcal{M}$ are allowed; thus, the combined action $S_{EH} + S_{GHY}$ produces the right motion equations even in the presence of a manifold with boundary. In (\ref{terGH}), $h$ is the determinant of the induced metric on the boundary surface, $K=\nabla_{\mu}n^{\mu}$ is the trace of the extrinsic curvature $K_{\mu\nu}$ of the boundary, and $\epsilon=+1\, (-1)$ depending on the timelike (spacelike) character of the boundary surface described by the unit normal $n^{\mu}$.

In a different realm, it was suggested in \cite{Oshita:2017nhn} that the formulation of GR relying on absolute parallelism, known as the teleparallel equivalent of GR (TEGR) \cite{TEGRbook,Manuel}, could naturally incorporate a sort of GHY counter-term due to its inherent first derivative action for the tetrad field $e^a(x^\mu)$. This suggestion is based on the Riemann-Weitzenb\"{o}ck link expressed by
\begin{equation}
  \label{ST}
R =T-2e^{-1}\partial_{\mu}(e \,T^{\mu})\,,
\end{equation}
where $e=det(e^a_{\mu})=\sqrt{g}$, and $T$ is the so called Weitzenb\"{o}ck scalar \footnote{An unfortunate denomination for an object which transforms as a scalar only under general coordinate changes and \emph{global} Lorentz transformations on the tangent plane.} 
\begin{equation}\label{elt}
    T =\frac{1}{4}T^a{}_{\mu\nu}T_a{}^{\mu\nu}-\frac{1}{2}T_{\mu\nu}{}^aT_a{}^{\mu\nu}
    -T_{\sigma\nu}{}^\sigma T_\mu{}^{\nu\mu}\,,
\end{equation}
which is constructed from the Weitzenb\"{o}ck torsion two-form $T^a=de^a$, having local spacetime components  
\begin{equation}\label{tamunu}
    T^a{}_{\mu\nu} = \partial_\mu e^a_\nu - \partial_\nu e^a_\mu\,. 
\end{equation}
We should emphasize that the tetrad $e^a_\nu$ carry Lorentz indexes $a$ ($(0), \dots ,(3)$, for specific components), as well as spacetime-type indexes $\mu$. The total derivative term in (\ref{ST}) is based on the ``vector" part
of the torsion, defined by the contraction $T^{\mu}=T^{\alpha\mu}_{\,\,\,\,\,\,\alpha}$. The interpretation of this total derivative term as the GHY term is then plausible, except for a crucial reason: the GHY term is invariant under local Lorentz transformations (LLT), because is defined in terms of the metric $\bold{g}=e^a\,e^b\,\eta_{ab}$, which is insensitive to local changes of the sort $e^a\rightarrow e^{a\,\prime}=\Lambda^{a\,\prime}_{\,\,b}\,e^b$. This means that the contribution of the GHY term to the EH action is independent of the chosen frame, as it should be. In stark contrast, the total derivative constructed from $T^{\mu}$ is not local Lorentz invariant; it is defined in terms of an object which does not transform as a vector under local Lorentz changes, the ``vector" part of the torsion tensor. At the end, if we are looking for a relation between the two boundary terms, it is then obvious that a kind of gauge fixing should apply. Currently, torsional boundary terms are being investigated not only from the perspective of the absolute parallelism (teleparallel) formulation of GR, but in extended torsion-based theories of gravity as well, see e.g. \cite{Martin}-\cite{Christian2}. The purpose of this work is to study this in more detail by paying attention to the emergence of preferred reference frames, even in the intrinsically Lorentz invariant framework provided by GR.

\section{Boundary terms}\label{sec}
The variational problem should be consistent with keeping the metric fixed on the boundary surface. If we incorporate this condition, the variation of the GHY term is 
\begin{equation}
  \label{vterGH}
 \delta S_{GHY} =\frac{\epsilon}{16 \pi }\oint_{\partial \mathcal{M}} d^3x \sqrt{h}\, n^{\mu}\,h^{\alpha\beta}\partial_{\mu}\delta g_{\alpha\beta}\,,
\end{equation}
where $h^{\alpha\beta}=g^{\alpha\beta}-\epsilon n^{\alpha}n^{\beta}$ is the induced metric on the boundary surface, and $h$ its determinant. The gauge fixing condition used in \cite{Oshita:2017nhn} involves the obvious implication
\begin{equation}
 \delta e^{a}_{\mu}\mid_{\partial \mathcal{M}}=0\Rightarrow \delta g_{\mu\nu}\mid_{\partial \mathcal{M}}=0\,. \label{equiv}
\end{equation}
If we then force to fix not only the metric but the tetrad field as well on the boundary, the term $h^{\alpha\beta}\partial_{\mu}\delta g_{\alpha\beta}$ written in terms of the tetrad reads
\begin{eqnarray}
 \nonumber    h^{\alpha\beta}\partial_{\mu}\delta g_{\alpha\beta}&=& h^{\alpha\beta}\partial_{\mu}(\delta e^a_{\,\,\alpha}e^b_{\,\,\beta}+e^a_{\,\,\alpha}\delta e^b_{\,\,\beta})\eta_{ab}\\
  &=&2e_a^{\,\,\alpha}(\partial_{\mu}\delta e^a_{\,\,\alpha}-\partial_{\alpha}\delta e^a_{\,\,\mu})\,.   \label{terGH2}
\end{eqnarray}
Here, we have used $g_{\alpha\beta}=\eta_{ab}e^{a}_\alpha e^{b}_\beta$ and  $e_{a}^{\,\,\mu}=h_a^{\,\,\mu}+\epsilon\, n_a n^\mu$, with $n_a=e_a^{\,\,\mu}n_\mu$ (here $e_{a}^{\,\,\mu}$ are the components of the inverse tetrad). Take note that only terms contracted in the direction of the unit normal $n^{\mu}$ appear. It is clear now that, due to the Dirichlet-type condition $\delta e^a\mid_{\partial \mathcal{M}}=0$, the expression (\ref{terGH2}) does not transform anymore as a covariant vector under the action of a LLT on the tetrad field. 

On the other hand, and for the sake of comparison, let us consider the variation of the torsional surface term coming from the total derivative in (\ref{ST}), namely
\begin{equation}
  \label{tertel}
 \delta S_{B} =\frac{\epsilon}{8 \pi }\oint_{\partial \mathcal{M}} d^3x \sqrt{h}\,n_{\mu}\delta T^{\alpha\mu}_{\,\,\,\,\,\,\alpha}\,,
\end{equation}
because $\delta n^{\mu}=0$. Then, by considering $T^{\alpha\mu}_{\,\,\,\,\,\,\alpha}=e_{a}^{\,\,\alpha}(\partial ^{\mu} e^{a}_{\,\,\alpha}-\partial _{\alpha} e^{a\mu})$, the variation $\delta T^{\alpha\mu}_{\,\,\,\,\,\,\alpha}$ results to be
\begin{equation}
  \label{tertel1}
 \delta T^{\alpha\mu}_{\,\,\,\,\,\,\alpha} =  e_{a}^{\,\,\alpha}(\partial^{\mu}\delta e^a_{\,\,\alpha}-\partial_{\alpha}\delta e^a_{\,\,\mu})\,.
\end{equation}
It is clear that the condition $\delta e^a\mid_{\partial \mathcal{M}}=0$ leads to the remarkable identity pointed out in \cite{Oshita:2017nhn}, i.e.   
\begin{equation}
  \label{igterb}
 \delta S_{GHY}=\delta S_{B}\,.
\end{equation}
Even though (\ref{igterb}) is certainly true at this level, we find this approach somewhat inconvenient for at least three reasons:
\begin{enumerate}
    \item Due to the non-local Lorentz invariant character of the equation (\ref{igterb}), the values of $S_{GHY}$ and $S_{B}$ depend on the chosen frame. If a concrete physical meaning is attributed to such quantities, as will be the case in a moment, we should explain why their values are frame-dependent, and what consequences this dependence has from a physical point of view.

    \item As we will see, the computation of $S_{GHY}$ and $S_{B}$ involves almost inevitable background subtractions, because they turn divergent as the boundary goes to infinity. Although background states concerning the metric can be usually defined, at least for spaces having a proper asymptotic behavior, it is by no means clear what the asymptotic form of the tetrad should be for those well defined metric backgrounds. 
    
    \item Finally, and from a slightly more conceptual point of view, we should recall that GR is a theory governing the metric field; it does not seem clear why one should fix the tetrad field on the boundary surface. Actually, the converse implication in (\ref{equiv}) is not true in general, because keeping the metric fixed on the boundary just means that the tensor $e^{a}_{\mu}e^{b}_{\nu}\eta_{ab}$ is held fixed, not $e^{a}_{\mu}$ itself. Any Lorentz transformed tetrad $e^{a'}_{\mu}=\Lambda^{a'}_{b}e^{b}_{\mu}$ will leave the metric invariant and, in particular, its variation on the boundary. 
\end{enumerate}

\bigskip
The importance of the GHY term in what concerns to black hole thermodynamics is a well established fact since the late seventies. Actually, the evaluation of that term is associated to the spacetime free energy divided by the temperature in the path integral approach. It is found that the entropy results
\begin{equation}
\label{thermo}
    \mathcal{S}=\beta <\mathcal{E}>-S_{GHY}\,,
\end{equation}
where $\beta=8\pi M$ is the period of the (identified) imaginary time, which equals the inverse of the temperature, and $<\mathcal{E}>$ is the expectation value of the energy, see \cite{Rengo2} and \cite{Rengo3} for a detailed treatment. In the following we shall expand on the points (1)-(3) raised above.

\section{Entropy, energy, and the background}

In order to illustrate the gauge-fixing-torsional approach to the spacetime free energy, let us consider the spherically symmetric, static and asymptotically flat \emph{Euclidean} line element written in Schwarzschild-like coordinates $(\tau=i t,r,\theta,\phi)$
\begin{equation}
  \label{metsph}
ds^2=f\,d\tau^2+f^{-1}\,dr^2+r^2d\Omega^2\,,
\end{equation}
where $f=f(r)\rightarrow 1$ as $r\rightarrow\infty$. Adapted to (\ref{metsph}), there are an infinite number of tetrads $e^a_{\,\,\mu}$ verifying $\tilde{g}_{\mu\nu}=\tilde{\eta}_{ab}e^a_{\,\,\mu}\,e^b_{\,\,\nu}$, where tilde refers to Euclidean (Wick-rotated, i.e., $t\rightarrow \tau=i t$ transformed) objects. Worth of noticing is the fact that the tetrad can be considered as an intrinsic Euclidean frame of fields by itself, having signature $+,+,+,+$ from the onset. Actually, the pseudo-Riemannian character of the usual metric field $g_{\mu\nu}=\eta_{ab}\,e^a_{\,\,\mu}e^b_{\,\,\nu}$ is produced by the standard pseudo-Riemannian Minkowski metric $\eta_{ab}=diag(-1,1,1,1)$, and not by the signature of the tetrad itself. In this way, Wick-rotated metrics are obtained from the tetrad field by Wick-rotating the Minkowski metric alone.

The Euclidean, torsional boundary action $\tilde{S}_{B}$ coming from integrating the term $e\, T$ in (\ref{ST}) is
\begin{equation}\label{intbount}
     \tilde{S}_B=\frac{1}{8\pi}\int_{\mathcal{M}}\,\partial_{\mu}(\tilde{e}\, \tilde{T}^{\alpha\mu}_{\,\,\,\,\,\,\,\, \alpha})\, d^4x=\frac{1}{8\pi}\oint_{\partial\mathcal{M}}\sqrt{\tilde{h}}\, \tilde{n}^{\mu} \,\tilde{T}^{\alpha}_{\,\,\,\,\mu \alpha}d^3x,
\end{equation}
where $\tilde{h}=det(\tilde{h}_{\mu\nu})$, and $\tilde{h}_{\mu\nu}=\tilde{g}_{\mu\nu}-\epsilon\, \tilde{n}_{\mu}\tilde{n}_{\nu}$. In what follows we will consider a radial normal vector $n^{\mu}=\sqrt{f}\,(0,1,0,0)$, which is unaffected by the Wick rotation, reason why $\tilde{h}_{\mu\nu}=\text{diag}(f,0,r^2,r^2\sin^2\theta)$ and $\tilde{h}=f\,r^4\,\sin^2\theta$. We can proceed now to explicitly compute (\ref{intbount}), thus

\begin{equation}
    \tilde{S}_B=\frac{1}{8\pi}\int_0^{\beta}d\tau \int_0^{2\pi}d\phi\int_0^{\pi}d\theta\,\sin\theta \,f \,r^2 \,\, \tilde{T}^{\alpha}_{\,\,\,\,\,r \alpha}\,,
\end{equation}
or simply  
\begin{equation}\label{else}
    \tilde{S}_B=-\frac{\beta}{2}r^2 f\,\tilde{T}^{\alpha}_{\,\,\,\,\,\alpha r}\,,
\end{equation}
where we used $\tilde{T}^{\alpha}_{\,\,\,\,\,r \alpha}=-\tilde{T}^{\alpha}_{\,\,\,\,\,\alpha r}$. Hence, we see that once the coordinates and metric are fixed as in (\ref{metsph}), the radial component $T_{r}=T^{\alpha}_{\,\,\,\alpha r}$ is still dependent on the chosen frame. Furthermore, due to the fact the radial component of the torsion vector is $T^{\alpha}_{\,\,\,\,r \alpha}=e_{a}^{\,\,\alpha}(\partial_{r} e^{a}_{\,\,\alpha}-\partial _{\alpha} e^{a}_{\,\,r})$, is not affected by the Wick rotation, i.e., $\tilde{T}^{\alpha}_{\,\,\,\,\,r \alpha}=T^{\alpha}_{\,\,\,\,\,r \alpha}$. In other words, the evaluation of (\ref{else}) and its subsequent interpretation as the spacetime free energy would be rather meaningless if one would be unable to confer a concrete physical meaning to the different tetrads associated to a given metric. With the purpose of illustrating this in detail, we will consider three different tetrads adapted to the metric (\ref{metsph}): 

\begin{itemize}
    \item The \emph{diagonal} tetrad  
    \end{itemize}
\begin{align}
e^{(0)}&=\sqrt{f}\, dt\,, \nonumber\\
e^{(1)}&=\frac{dr}{\sqrt{f}}\,,  \nonumber\\
e^{(2)}&=r\,d\theta\,,  \nonumber\\
e^{(3)}&=r\,\sin \theta \,d\phi\,. \label{diagtet} 
\end{align}  
This is actually the tetrad considered in \cite{Oshita:2017nhn}. It has no direct or natural physical meaning; hence, it cannot be interpreted as an observer in a given state of motion. The $r$-component of the vector torsion needed to compute (\ref{else}), results
\begin{equation}
    T^{\alpha}_{\,\,\,\alpha r}=\frac{2}{r}+\frac{f^{\prime}}{2f}\,.
\end{equation}
As a consequence of this, the boundary term (\ref{else}) in the Schwarzschild case $f=1-2M/r$ can be written as
\begin{equation}
    \tilde{S}_B=12\pi M^2-8\pi M r\,.
\end{equation}
It is evident that $S_B$ should be subjected to a background subtraction of one sort of another in the limit $r\rightarrow \infty$.
\begin{itemize}
    \item The \emph{static} frame 
    \end{itemize}
\begin{align}\label{staticf}
 e^{(0)}&=\sqrt{f}\, dt\,, \nonumber\\
  e^{(1)}&=\frac{\sin\theta \cos\phi\, dr}{\sqrt{f}} + r\, d\Omega^{(1)}\,,  \nonumber\\ 
  e^{(2)}&=\frac{\sin\theta \sin\phi\, dr}{\sqrt{f}} + r\, d\Omega^{(2)}\,,\nonumber\\
  e^{(3)}&=\frac{\cos\theta\, dr}{\sqrt{f}} - r \sin\theta\, d\theta\,,
\end{align}
where
\begin{align}\label{Schframedef}
d\Omega^{(1)}&=\cos \theta  \cos \phi \, d\theta - \sin \theta  \sin \phi \, d\phi\,,\nonumber\\ 
d\Omega^{(2)}&=\cos \theta  \sin \phi \, d\theta + \sin \theta  \cos \phi \, d\phi\,. \nonumber 
\end{align}
This frame is the one adapted to an observer at rest with respect to the black hole, see \cite{Maluf1} and \cite{Maluf2}. In this case, the $r$-component of the vector torsion is
\begin{equation}\label{vectorestatic}
    T^{\alpha}_{\,\,\,\alpha r}=\frac{2}{r}+\frac{f^{\prime}}{2f}-\frac{2}{r\sqrt{f}}\,,
\end{equation}
and the boundary term for $f=1-2M/r$ is now
\begin{equation}
    \tilde{S}_B=4\pi M r\left(2\sqrt{1-2M/r}+3M/r-2\right) \,.
\end{equation}
In the limit $r\rightarrow \infty$, we have
\begin{equation}
\tilde{S}_B=4\pi M^2 \,,
\end{equation}
which means that this frame has a proper asymptotic behavior (see comments below). 
\begin{itemize}
    \item The \emph{free falling} frame
\end{itemize}
\begin{align}\label{Schframe}
 e^{(0)}&=- dt-\frac{\sqrt{1-f}}{f}\, dr\,, \nonumber\\
  e^{(1)}&=\sin \theta  \cos \phi \left(\sqrt{1-f} \, dt  + \frac{dr }{f}\right)+r\, d\Omega^{(1)}\,,\nonumber\\ 
  e^{(2)}&=\sin \theta  \sin \phi \left(\sqrt{1-f} \, dt  + \frac{dr }{f}\right)+r\, d\Omega^{(2)}\,,\nonumber\\ 
 e^{(3)}&=\sqrt{1-f}\cos \theta  \, dt  + \frac{\cos \theta  }{f}\, dr-r \sin \theta  \, d\theta\,. 
\end{align}
This is the tetrad adapted to a non rotating observer in radial free fall \cite{Maluf2}. This frame belongs to the equivalence class of frames giving $T=0$ for the Schwarzschild spacetime, see \cite{FYF}. Now, the $r$-component of the vector torsion is 
\begin{equation}\label{compvector}
    T^{\alpha}_{\,\,\,\alpha r}=\frac{-4+4f+rf^{\prime}}{2\,rf}\,,
\end{equation}
which, for $f=1-2M/r$ ends up being 
\begin{equation}\label{vectorsch}
    T^{\alpha}_{\,\,\,\alpha r}=\frac{3M}{2Mr-r^2}\,,
\end{equation}
and the boundary term (\ref{else}), can be written as 
\begin{equation}
\label{BTSF}
    \tilde{S}_B= 12\pi M^2\,.
\end{equation}
Again, this frame does not require any background subtraction.

Using standard path-integral arguments, and identifying the expectation value of the energy of the Schwarzschild space with the black hole mass $<\mathcal{E}>=M$, we have the different entropy values coming from Eq. (\ref{thermo}) with $S_{GHY}=S_{B}$:
\begin{align}
\mathcal{S}&=-4\pi M^2+8\pi M r=-\frac{A}{4}+\frac{Ar}{2}\,,\,\,\,\text{(diagonal)}\label{endiagf}\\
\mathcal{S}&=4\pi M^2=\frac{A}{4}\,,\,\,\,\,\,\,\,\,\,\,\,\,\,\,\,\,\,\,\,\,\,\,\,\,\,\,\,\,\,\,\,\,\,\,\,\,\,\,\,\,\,\,\,\,\,\,\,\,\,\,\,\,\text{(static)}\label{enestf}\\
\mathcal{S}&=-4\pi M^2=-\frac{A}{4}\, ,\,\,\,\,\,\,\,\,\,\,\,\,\,\,\,\,\,\,\,\,\,\,\,\,\,\,\,\,\,\,\,\,\,\,\,\,\,\,\,\,\,\,\text{(free falling)}\label{enschf}
\end{align}
where $A=16\pi M^2$ is the area of the Schwarzschild horizon. We thus think that this variety and disparity of results is enough for considering a more systematic approach to the black hole thermodynamics within this context, illustrating in this way the point (1) raised above. 

\bigskip

We now discuss point (2). Asymptotically flat spacetimes having a metric of the form
$g_{\mu\nu}=\text{diag}(-f,f^{-1},r^2,r^2\sin^2\theta)$ require $f\rightarrow 1$ as $r\rightarrow\infty$. The reason why some tetrads are so badly behaved asymptotically, as in the case of the diagonal one leading to (\ref{endiagf}), relies in the asymptotic structure of the torsion components they produce. Although the frame (\ref{diagtet}) is consistent with the asymptotically flat metric just mentioned, it produces the following torsion (spacetime) components at infinity, 
\begin{align}\label{staticfas}
 T^t_{\,\, tr}&= f^{\prime}_{\infty}/2\,,\,\,\,\,\,\,\,\,\, T^{\theta}_{\,\, r \theta}=-1/r\,,  \nonumber\\
 T^{\phi}_{\,\, r \phi}&= -1/r\,,\,\,\,\,\,T^{\phi}_{\,\, \theta \phi}= - \cos \phi / \sin \theta\,.  
 \end{align}
While some components vanish as $1/r$ when $r\rightarrow\infty$, we see that others simply become functions of the angular coordinates. It would seem rather strange, if not wrong, to be in the need of performing a background subtraction on the basis of a tetrad field which produces non-null torsion components at spatial infinity, specially if other tetrads leading to the same metric tensor produce null results in the first place. In turn, the asymptotic behavior of both the static and free falling frames (\ref{staticf}) and (\ref{Schframe}), is just  
\begin{align}\label{staticfas}
 e^{(0)}_{\infty}&= dt\,, \nonumber\\
  e^{(1)}_{\infty}&=\sin\theta \cos\phi\, dr + r\, d\Omega^{(1)}\,,  \nonumber\\
  e^{(2)}_{\infty}&=\sin\theta \sin\phi\, dr + r\, d\Omega^{(2)}\,, \nonumber\\
  e^{(3)}_{\infty}&=\cos\theta\, dr - r \sin\theta\, d\theta\,.
\end{align}
To be precise, the associated asymptotic torsion components for the static frame (\ref{staticf}) are
\begin{equation}\label{toricfas}
 T^{t}_{\,\,tr} = \frac{f^{\prime}_{\infty}}{2}\,,\,\,\,\,\, T^{\theta}_{\,\,\,  r \theta}=(-1+f^{-1/2}_{\infty})/r \,,\,\,\,\,\,  
 T^{\phi}_{\,\,\,  r \phi} =(-1+f^{-1/2}_{\infty})/r\,,
 \end{equation}
which fall off as $1/r^2$ for the Schwarzschild function $f=1-2M/r$. Actually, this is not a coincidence, but a consequence of the fact that (\ref{staticfas}) is the Euclidean frame $e^{a}=\delta^a_{\,\,\,b}\,dx^b$ -- $x^a$ stands for ($t, x^{i}$), where $x^{i}$ are cartesian coordinates-- written in spherical coordinates. Because of that it has null torsion, unlike the asymptotic diagonal frame $e^a=\text{diag}(1,1,r,r\sin\theta)$ associated to Minkowski line element written in spherical coordinates. Then, it seems that the present formalism not only produces sensibly different outcomes depending on the choice of frame, but also involves a rather arbitrary, physically undriven choice of background in some cases.

However, despite the various drawbacks just established, it appears that every now and then the right results come out. For instance, as was shown in Eq. (\ref{enestf}), the static frame (\ref{staticf}) does reproduce the correct black hole entropy without any background subtraction. In the following lines we will show that the correct energy of the gravitational field, i.e. $<\mathcal{E}>=M$, can be actually obtained from this frame as a consequence of the well known teleparallel energy-momentum pseudo current \cite{Bras}, see also \cite{Maluf1} and \cite{Maluf2}. Let us remember that by starting from the Einstein equations written in its teleparallel form
\begin{equation}
e^{-1} \partial _{\mu}(e \, S^{a\mu}_{\,\,\,\,\,\,\lambda})-T_{\rho\mu}^{\,\,\,\,\,\,a}S^{\rho \mu}_{\,\,\,\,\,\,\lambda} + \frac 14 e^a_{\,\,\lambda} T = 4\pi \,T^a_{\,\,\lambda}\,,
\label{movetrg}
\end{equation}
and defining
\begin{equation}
j^a_{\,\,\lambda}=\frac{1}{4\pi}(T_{\rho\mu}^{\,\,\,\,\,\,a}S^{\rho \mu}_{\,\,\,\,\,\,\lambda} - \frac 14 e^a_{\,\,\lambda} T)\,,
\label{corriente}
\end{equation}
the motion equations (\ref{movetrg}) can be cast into the form
\begin{equation}
e^{-1} \partial _{\mu}(e \, S^{a\mu}_{\,\,\,\,\,\,\lambda})-4\pi j^a_{\,\,\lambda} = 4\pi \,T^a_{\,\,\lambda}\,,
\end{equation}
where $T^a_{\,\,\lambda}$ stands for the energy-momentum tensor of the matter fields. In all these expressions the \emph{superpotential} $S^{a\mu}_{\,\,\,\,\,\,\lambda}$ plays a prominent role. It is defined in terms of the torsion as
\begin{align}
    S^a{}_{\mu\nu} = \frac{1}{4}(T^a{}_{\mu\nu}-T_{\mu\nu}{}^a+T_{\nu\mu}{}^a) +
    \frac{1}{2}(\delta^a_\mu T_{\sigma\nu}{}^\sigma-\delta^a_\nu T_{\sigma\mu}{}^\sigma)\,,\nonumber
 \end{align}
and it verifies $T=S^{a\mu}_{\,\,\,\,\,\,\lambda}T_{a\mu}^{\,\,\,\,\,\,\lambda}$. Due to the skew-symmetry of $S^{a\mu}_{\,\,\,\,\,\,\lambda}$ with respect to the last two indexes, a natural conserved current appears
\begin{equation}
\partial ^{\lambda}[e \,J^a_{\,\,\lambda}]=0\,,\,\,\,\,\,\,J^a_{\,\,\lambda}=j^a_{\,\,\lambda}+ T^a_{\,\,\lambda}\,.
\end{equation}
In this way, $j^a_{\,\,\lambda}$ is viewed as the energy-momentum current of the gravitational field. Because of that, we can define the total energy-momentum four-vector of the gravitational field by integrating the $\lambda=t$ component of $J^a_{\,\,\lambda}$ over a spatial hypersurface $V$ defined by $t=$constant (we shall use $J^{a\lambda}$ instead of  $J^a_{\,\,\lambda}$), namely 
\begin{equation}
p^a=\int _V d^3x \,e \,J^{a t}\,.
\label{vem}
\end{equation}
Using the motion equations, (\ref{vem}) can be written as:
\begin{equation}
p^a=\frac{1}{4\pi }\int _V d^3x \,\partial _{\mu}(e S^{a\mu t})=\frac{1}{4\pi }\oint _{S(V)} dS_{\mu}\, e\, S^{a\mu t}\,,
\label{vem2}
\end{equation}
so the total energy present in the gravitational and matter fields contained in the volume $V$ is just
\begin{equation}
\mathcal{E}=p^{(0)}=\frac{1}{4\pi }\oint _{S(V)} dS_{\mu}\, e\, S^{(0)\mu t}\,.
\label{vem3}
\end{equation}
It is quite important to stress once again that the energy-momentum four-vector so obtained, Eq. (\ref{vem2}), is not a true Lorentz vector, so its relevance as a genuine, pure gravitational current should be minimised. However, as mentioned countless times in the literature, it produces correct results in a number of physical situations. But these results should be properly contextualized and interpreted, because they strongly depend on the choice of frames. What follows is a tentative to further clarify the role of the static frame (\ref{staticf}) as a candidate to fully describe the thermodynamics of the Schwarzschild black hole. 

Due to the geometry under consideration, we are interested in the components $S^{(0)r t}$; remember that the normal vector is radial, so $\mu=r$. We start from
\begin{equation}
    S^{b \mu \nu}=g^{\rho \mu}g^{\pi \nu} \eta^{ab} e^{\,\,\sigma}_{a}S_{\sigma \rho \pi}\,. 
\end{equation}
Then we have 
\begin{equation}
    S^{(0) r t}= \eta^{(0)(0)} g^{tt}g^{rr} (e^{t}_{\,\,(0)}S_{ttr}+e^{r}_{\,\,(0)}S_{rtr})\,, 
\end{equation}
or 
\begin{equation}\label{enden}
    S^{(0) r t}= e^{t}_{\,\,(0)}S_{ttr}+e^{r}_{\,\,(0)}S_{rtr}\,. 
\end{equation}
For the static frame (\ref{staticf}) we see that $e^{r}_{\,\,(0)}=0$. Moreover, we obtain
\begin{equation}\label{endenalgo}
 e^{t}_{\,\,(0)}=\frac{1}{\sqrt{f}}\,,\,\,\,\,  S_{ttr}=\frac{\sqrt{f}-f}{r}\,.
\end{equation}
At the end we have that, for $f=1-2M/r$, (\ref{enden}) is 
\begin{equation}\label{endenfin}
  S^{(0) r t}=\frac{1-\sqrt{1-2M/r}}{r}\approx\frac{M}{r^2},
\end{equation}
as $r\rightarrow\infty$. In this way, after integrating as in (\ref{vem3}), we see that the total gravitational energy is just $\mathcal{E}=M$. 

Before to conclude this section, we should mention once again that the physically reasonable result $\mathcal{E}=M$, is by no means universal. As shown in \cite{Maluf2}, the free falling frame (\ref{Schframe}) adapted to the Schwarzschild geometry conduces to $\mathcal{E}=0$, in agreement with the equivalence principle. This is very easy to check by noting that in this case we have
\begin{equation}
    e^{t}_{\,\,(0)}=-(1-2M/r)^{-1}\,,\,\,\,\, e^{r}_{\,\,(0)}=\sqrt{2M/r}\,.
\end{equation}
After routine calculations we find
\begin{equation}
    S_{ttr}=-\frac{2M}{r^2}\,,\,\,\,\, S_{rtr}=\frac{\sqrt{2M/r}}{2M-r}\,,
\end{equation}
which, at the end, makes Eq. (\ref{enden}) to be
\begin{equation}\label{enden2}
    S^{(0) r t}= 0\,,
\end{equation}
giving thus $\mathcal{E}=0$ in this frame.

\section{A suggestion for finding proper frames complying  $\delta e^{a}_{\mu}\mid_{\partial \mathcal{M}}=0$}

The facts just obtained, i.e., the finding of a tetrad leading to the right thermodynamic variables, lead us to ask if any other frame is also able to reproduce these nice results. It is actually possible to obtain a full equivalence class of frames by using the so called remnant symmetries \cite{grupo}. Plainly, these are local Lorentz transformations leaving invariant the Weitzenb\"{o}ck scalar $T$, i.e., they are local Lorentz changes under which $T$ becomes a true Lorentz scalar. This construction requires a seed tetrad to start from, and from which the whole equivalence class is built. Due to the link (\ref{ST}), transformations belonging to the remnant equivalence class will leave invariant the torsional surface term as well, assuring thus the constancy of the gravitational free energy. In this regard, the identification of the remnant symmetries associated to a given space is important because they provide a relatively systematic procedure to obtain sufficient --although not necessary-- conditions for exploring the invariance of the surface term.

However we should bear in mind that, at this level, the remnant symmetries in question must verify the gauge condition $\delta e^{a}_{\mu}\mid_{\partial \mathcal{M}}=0$. For instance, local rotations on the 2-sphere are not permitted, because they will generate non null variations of the tetrad on the boundary surface. This leaves us just with a radial boost acting on the static frame, this is
\begin{align}\label{staticfboo}
 e^{(0)}_{\nearrow}&=\sqrt{f}\,\cosh\Phi\,dt+\sinh\Phi\left(\frac{\sin\theta \cos\phi \,dr}{\sqrt{f}}+ r\,d\Omega^{(1)}\right) \,, \nonumber\\
  e^{(1)}_{\nearrow}&=\sqrt{f}\,\sinh\Phi\,dt+\cosh\Phi\left(\frac{\sin\theta \cos\phi\, dr}{\sqrt{f}} +r\,d\Omega^{(1)}\right)\,,  \nonumber\\ 
  e^{(2)}_{\nearrow}&=\frac{\sin\theta \sin\phi\, dr}{\sqrt{f}} + r\, d\Omega^{(2)}\,,\nonumber\\
  e^{(3)}_{\nearrow}&=\frac{\cos\theta\, dr}{\sqrt{f}} - r \sin\theta\, d\theta\,.
\end{align}
Here $\Phi=\Phi(r,\theta,\phi)$ is the local boost parameter. We found that the Weitzenb\"{o}ck scalar reads
\begin{equation}\label{eltconb}
T_{\nearrow}=\frac{2}{r^2}\,[1+f+r\,f^{\prime}-2\sqrt{f}-r\, (\sqrt{f}\,)^{\prime}\, ]\,,
\end{equation}
which is independent of the boost parameter $\Phi$, showing that this radial boost is actually a Lorentz transformation belonging to the remnant group associated to the static frame. In other words, the value of the free energy is insensitive to this local change of frame. In turn, the spacetime energy needs revision in the light of the boost performed to the tetrad. In allusion to the Eq. (\ref{enden}), after the boost we now obtain
\begin{align}\label{elsconb}
 e^{t}_{\nearrow\,(0)}&=\frac{\cosh\Phi}{\sqrt{f}}\,,\,\,\,\,\,\,\, S_{ttr}=\frac{\sqrt{f}-f}{r}\,,\nonumber\\
 e^{r}_{\nearrow\,(0)}&=-\frac{\sinh\Phi}{\sqrt{f}}\,,\,\, S_{rtr}=\frac{\cos\theta\cos\phi\,\Phi_{,\theta}-\csc\theta \sin\phi\,\Phi_{,\phi}}{2 r\sqrt{f}}\,.
\end{align}
Therefore we see that the asymptotic behavior $S^{(0)rt}\approx\frac{M}{r^2}$ as $r\rightarrow\infty$ for $f=1-2M/r$ is obtained provided $\Phi(r,\theta,\phi)\rightarrow 0$ in the same limit. This asymptotic requirement for $\Phi$ is somewhat expected in the light of the asymptotic form of the frame after the boost; as a matter of fact, the $r\rightarrow\infty$ limit of (\ref{staticfboo}) requires $\Phi\rightarrow 0$ in order for the frame to preserve its asymptotic Euclidean character. Notice, however, that $T=S^{a\mu}_{\,\,\,\,\,\,\lambda}T_{a\mu}^{\,\,\,\,\,\,\lambda}$ is not boost-dependent, see Eq. (\ref{eltconb}); this means that the changes of the components of $S^{a\mu}_{\,\,\,\,\,\,\lambda}$ and $T_{a\mu}^{\,\,\,\,\,\,\lambda}$ cancel out. Due to the fact that the gravitational energy depends solely on the components of $S^{a\mu}_{\,\,\,\,\,\,\lambda}$, the $\Phi\rightarrow 0$ requirement as $r\rightarrow\infty$ is a sufficient condition to preserve its value. In the language developed in \cite{Rubilar}, this boost is an \emph{asymptotically global} Lorentz transformation; under these changes, the pseudo-vector $p^a$ given in Eq. (\ref{vem2}) becomes a true Lorentz vector. Hence, at the end, there is an infinite number of observers perceiving the correct black hole thermodynamics; these are characterized by the tetrads $e^{a}_{\nearrow}$ given in Eq. (\ref{staticfboo}) with the sole imposition of having $\Phi\rightarrow 0$ as $r\rightarrow\infty$. In the next section we show that there are really many more.

\section{Free tetrad variations on the boundary surface}

What remains of the article is to discuss point (3) addressed in section \ref{sec}. With this purpose in mind let us analyze the variations of the boundary terms when the tetrad is not forced to be fixed on the boundary surface. If we focus first on the GHY boundary term, the integrand in (\ref{vterGH}) now reads
\begin{eqnarray}
 \nonumber    h^{\alpha\beta}\partial_{\mu}\delta g_{\alpha\beta}&=& h^{\alpha\beta}\partial_{\mu}(\delta e^a_{\,\,\alpha}e^b_{\,\,\beta}+e^a_{\,\,\alpha}\delta e^b_{\,\,\beta})\eta_{ab}\,,\\
  &=& \text{terms}\, (\partial_{\mu}\delta e)+ \text{terms}\, (\delta e)\,, \label{terGH2free}
\end{eqnarray}
being the two terms
\begin{eqnarray}
 \nonumber \text{terms}\, (\partial_{\mu}\delta e)&=&2e_a^{\,\,\alpha}(\partial_{\mu}\delta e^a_{\,\,\alpha}-\partial_{\alpha}\delta e^a_{\,\,\mu}+h_{\alpha}^{\,\,\beta}\partial_{\beta}\delta e^{a}_{\,\,\mu})\,, \\
   \text{terms}\, (\delta e)&=&h^{\alpha\beta}(\delta e_{b\alpha}\partial_{\mu}e^{b}_{\,\,\beta}+\delta e^b_{\,\,\beta}\partial_{\mu}e_{b\alpha})\,. \label{terGH4free}
\end{eqnarray}
Eq. (\ref{terGH2free}) with the definitions  (\ref{terGH4free}) is the extension of the Eq. (\ref{terGH2}). Now, in order to be operative, let us consider variations on the boundary surface having the structure
\begin{equation}
 \delta e^{a}_{\mu} \equiv e'^{a}_{\,\,\, \mu} - e ^{a}_{\,\,\, \mu}=\Lambda^{a}_{\,\,\,b}e^{b}_{\,\,\,\mu}-e^{a}_{\,\,\,\mu}\,.\label{vari}
\end{equation}
This seems reasonable because, once the metric and coordinates are fixed, two different tetrads on the boundary will be connected by a LLT. Thus, 
the variation of GHY boundary term gives
\begin{align}
\label{GHYV}
  \nonumber  \delta S_{GHY}&=2n^{\mu}e_{a}^{\,\,\alpha}[\Lambda^{a}_{\,\,\,b,\mu}e ^{b}_{\,\,\, \alpha}-\Lambda^{a}_{\,\,\,b,\alpha}e ^{b}_{\,\,\, \mu} +\Lambda^{a}_{\,\,\,b}(e ^{b}_{\,\,\, \alpha,\mu}-e ^{b}_{\,\,\, \mu,\alpha }) +e ^{a}_{\,\,\, \mu,\alpha} -e ^{a}_{\,\,\, \alpha,\mu}
  \nonumber +h_{\alpha}^{\,\,\beta}\partial_{\beta}\delta e ^{a}_{\,\,\, \mu}]\\
 & +n^{\mu}h^{\alpha\beta}[\delta e_{b\alpha}\partial_{\mu}e ^{b}_{\,\,\, \beta}+\partial_{\mu}e_{b\alpha}\delta e^{b}_{\,\,\, \beta}]\,.
\end{align}
On the other hand, the variation of the torsional boundary term (\ref{tertel}) results
\begin{equation}
  \label{tertelfree}
 \delta(T^{\alpha\mu}_{\,\,\,\,\,\,\alpha}) = \text{terms}\, (\partial_{\mu}\delta e)+ \text{terms}\, (\delta e)\,,
\end{equation}
where now we have
 \begin{eqnarray}
  \label{tertel2free}
\nonumber  \text{terms}\, (\partial_{\mu}\delta e)&=&e_a^{\,\,\alpha}(\partial^{\mu}\delta e^a_{\,\,\alpha}-\partial_{\alpha}\delta e^{a \mu})\,, \\
   \text{terms}\, (\delta e)&=&\delta e_{a}^{\,\,\alpha}(\partial^{\mu}e^a_{\,\,\alpha}-\partial_{\alpha}e^{a\mu})\,.
\end{eqnarray}
This last result is the generalization of (\ref{tertel1}). Hence, considering Eq. (\ref{vari}), the variation of the torsional boundary term gives 
\begin{eqnarray}
\label{torsionalV}
 \nonumber   \delta S_{B}&=&n^{\mu}e_{a}^{\,\,\alpha}[\Lambda^{a}_{\,\,\,b,\mu}e ^{b}_{\,\,\, \alpha}  -\Lambda^{a}_{\,\,\,b,\alpha}e ^{b}_{\,\,\, \mu} +\Lambda^{a}_{\,\,\,b}(e ^{b}_{\,\,\, \alpha,\mu}-e ^{b}_{\,\,\, \mu,\alpha })
 +2(e ^{a}_{\,\,\, \mu,\alpha}-e ^{a}_{\,\,\, \alpha,\mu})]\\
  &&+ n^{\mu}(e^a_{\,\,\alpha,\mu}-e^a_{\,\,\mu,\alpha})\Lambda_{ab}e^{b\alpha}\,.
\end{eqnarray}
After equating the two contributions coming from (\ref{GHYV}) and (\ref{torsionalV}), and using (\ref{vari}), a relatively involved calculation gives us
\begin{equation}
n^{\mu}M_{\mu}=0\,, \label{cond}
\end{equation}
where the vector $M_{\mu}$ is
\begin{eqnarray}
 \nonumber && M_{\mu} =  e_a^{\,\,\,\alpha}\Big[ \Lambda ^{a}_{\,\,\,b,\mu}e^b_{\,\,\,\alpha}+  \Lambda ^{a}_{\,\,\,b,\alpha}e^b_{\,\,\,\mu}-2\epsilon\, n_{\alpha}n^{\beta}\Lambda ^{a}_{\,\,\,b,\beta}e^b_{\,\,\,\mu}+\\
 &&
    \Big(\Lambda ^{a}_{\,\,\,b}+(\Lambda^{-1})^{a}_{\,\,\,b}-2\delta^{a}_{\,\,\,b}\Big)\Big(e^b_{\,\,\,\alpha,\mu}+e^b_{\,\,\,\mu,\alpha}-2\epsilon\, n_{\alpha}n^{\beta}e^b_{\,\,\,\mu,\beta}\Big)\Big]\,.\label{elvecmu}
\end{eqnarray}

Equation (\ref{cond}) determines the set of LLT $\Lambda ^{a}_{\,\,\,b}(x^{\mu})$ allowed at the boundary $\partial\mathcal{M}$ in order to make $\delta S_{B}=\delta S_{GHY}$ for a given seed tetrad $e^a$. This set generates variations $\delta e^{a}$ on the surface boundary which are consistent with $\delta g_{\mu\nu}|_{\partial \mathcal{M}}=0$. Take note that $\Lambda ^{a}_{\,\,\,b}=\delta^{a}_{\,\,\,b}$ automatically solves Eq. (\ref{cond}), as it should be, because this implies $\delta e^a_{\mu}=0$ which is the case studied before.

We can obtain a further clue on the structure of (\ref{elvecmu}) by considering infinitesimal Lorentz transformations $\Lambda ^{a}_{\,\,\,b} = \delta ^{a}_{\,\,\,b}+\omega ^{a}_{\,\,\,b} + \mathcal{O} (\omega^2)$ with $\omega^a_{\,\,\, b} \ll 1$. In this case (\ref{elvecmu}) is
\begin{align}
\label{ILT}
 \nonumber M_{\mu}& =  e_a^{\,\,\,\alpha}\Big( \omega ^{a}_{\,\,\,b,\mu}e^b_{\,\,\,\alpha}+  \omega ^{a}_{\,\,\,b,\alpha}e^b_{\,\,\,\mu}-2\epsilon \,n_{\alpha}n^{\beta}\omega ^{a}_{\,\,\,b,\beta}e^b_{\,\,\,\mu}\Big)\\
 &=  \omega ^{a}_{\,\,\,a,\mu}+  e_a^{\,\,\,\alpha} e^b_{\,\,\,\mu}\Big(\omega ^{a}_{\,\,\,b,\alpha}-2\epsilon \,n_{\alpha}n^{\beta}\omega ^{a}_{\,\,\,b,\beta}\Big).\nonumber
\end{align}
Using the fact that $\omega _{ab}$ is anti-symmetric, the above equation gives
\begin{equation}
\label{ILT2}
 M_{\mu} =  e_a^{\,\,\,\alpha} e^b_{\,\,\,\mu}\Big(\omega ^{a}_{\,\,\,b,\alpha}-2\epsilon\, n_{\alpha}n^{\beta}\omega ^{a}_{\,\,\,b,\beta}\Big).
\end{equation}
The equation $n^{\mu}M_{\mu}=0$ with $M_{\mu}$ as in (\ref{ILT2}) defines the \emph{functional group of variations}. It consist on the small variations $\delta e^a_\mu=\omega ^{a}_{\,\,\,b}e^b_\mu$ on the boundary which lead to $\delta S_{B}=\delta S_{GHY}$. The action of the infinitesimal Lorentz matrices belonging to the functional group generates tetrads which are smoothly connected to the seed tetrad $e^a_\mu$ and which preserve the value of the thermodynamic quantities associated to the torsional boundary term.  

In order to find other frames leading to the right thermodynamic variables, the quest for remnant symmetries is still an important matter because they will automatically guarantee the invariance of the torsional boundary term. However, once we find other LLT producing variations of the tetrad on the surface boundary, we need to further check if they actually comply with Eq. (\ref{cond}), otherwise they will produce variations of the tetrad which are not consistent with the right motion equations. We proceed now to illustrate this subtle point.

Due to the fact that we are free, in principle, to locally rotate the tetrad on the boundary, let us consider as an example the following simple rotation
\begin{equation}  \label{rot}
\Lambda=\left(
\begin{array}{ccc}
\,\,\mathbb{I}\,\,\,\,\,\,\vline &\,\,\,\,\, \,\,\,\,\,\,\,\mathbb{O}  \\ \hline
\,\,\,\,\,\,\,\,\,\,\,\vline &\cos{\Theta} &\sin{\Theta} \\
\,\, \mathbb{O}\,\,\,\,\vline &-\sin{\Theta} &\cos{\Theta} \\
\end{array}
\right),
\end{equation}
where $\Theta(r,\theta,\phi)$ is the local parameter of the rotation, and $\mathbb{I},\mathbb{O}$ are the 2$\times$2 identity and null matrices respectively. The action of this rotation on the static frame (\ref{staticf}) gives
\begin{align}\label{staticfrot}
 e^{(0)}_{\circ}&=\sqrt{f}\, dt\,, \nonumber\\
  e^{(1)}_{\circ}&=\frac{\sin\theta \cos\phi\, dr}{\sqrt{f}} + r\, d\Omega^{(1)}\,,  \nonumber\\ 
  e^{(2)}_{\circ}&=\frac{\sin\theta \sin\phi \cos\Theta+\cos\theta\sin\Theta}{\sqrt{f}}\, dr + r\, d\Theta^{(1)}\,,\nonumber\\
  e^{(3)}_{\circ}&=\frac{\cos\theta\cos\Theta-\sin\theta\sin\phi\sin\Theta}{\sqrt{f}}\, dr - r\, d\Theta^{(2)}\,,
\end{align}
where
\begin{align}\label{nuedef}
 d\Theta^{(1)}&=\cos\Theta\sin\theta[(\cot\theta\sin\phi-\tan\Theta)\,d\theta+\cos\phi\, d\phi]\,, \nonumber\\
d\Theta^{(2)}&=\sin\Theta\sin\theta[(\cot\Theta+\cot\theta\sin\phi)\,d\theta+\cos\phi\, d\phi]\,.  \nonumber 
\end{align}
To know what kind of rotations of the form (\ref{rot}) are remnant symmetries of the static frame, we need to compute the Weitzenb\"{o}ck scalar $T$ for the rotated tetrad (\ref{staticfrot}). After doing so we find
\begin{equation}\label{eltconbrot}
T_{\circ}=\frac{2}{r^2}\,[1+f+r\,f^{\prime}+F-(2+F)(\sqrt{f}+r\, (\sqrt{f}\,)^{\prime}\,) ]\,,
\end{equation}
where the function $F$ is $F(r,\theta,\phi)=\cos\phi\cot\theta\,\Theta_{,\phi}+\sin\phi\,\Theta_{,\theta}$. A direct comparison between (\ref{eltconb}) and (\ref{eltconbrot}) leads us to conclude that $\Lambda$ is a remnant symmetry provided $F=0$; this differential equation for $\Theta(r,\theta,\phi)$ is solved by 
\begin{equation}\label{lsefesol}
\Theta(r,\theta,\phi)=\Theta_{0}\, R(r)\,\cos\phi\sin\theta\,,
\end{equation}
where $\Theta_{0}$ is an integration constant and $R(r)$ is a free function of the radial coordinate alone. Notice, however, that the asymptotic Euclidean structure of the the static frame (Eq. (\ref{staticfas})), is only preserved provided $R(r) \rightarrow 0$ when $r \rightarrow \infty$. Hereafter, we will then adopt this condition for the function $R(r)$.

The rotation (\ref{lsefesol}) also leaves invariant the gravitational energy. As a matter of fact, given the block structure of the tetrad (\ref{staticfrot}), we have $e^{r}_{\circ\,(0)}=0$ and
\begin{equation}\label{elsconrot}
 e^{t}_{\circ\,(0)}=\frac{1}{\sqrt{f}}\,,\,\,\,\,\,\,\, S_{ttr}=\frac{\sqrt{f}}{r}\,(2-2\sqrt{f}+F\,)\,.\nonumber
\end{equation}
In virtue that (\ref{lsefesol}) implies $F=0$, we obtain (\ref{endenalgo}) again. Then  $S^{(0) r t}$ turns out to be as in (\ref{endenfin}), reason why the total gravitational energy (\ref{vem3}) ends up being $\mathcal{E}=M$ in the Schwarzschild case.

Of course, the remnant symmetry just obtained assures the invariance of the torsional boundary term. It is instructive nonetheless, to check that the rotated frame is actually a solution of Eq. (\ref{cond}), namely, that it belongs to the class of frames smoothly connected to the static frame which produce variations on the boundary such that $\delta S_{B}=\delta S_{GHY}$. We will check this in the infinitesimal case (\ref{ILT2}), so $\Theta \ll 1$, $\cos\Theta\approx1$ and $\sin\Theta\approx\Theta$ in (\ref{rot}), which can be accomplished by requiring $R(r) \ll 1$ for large values of $r$. In this way we obtain $n^{\mu}M_{\mu}= F/r$, therefore $n^{\mu}M_{\mu}=0$ for $\Theta (r, \theta, \phi)$ as in (\ref{lsefesol}).


\section{Final comments}

Along these lines we tended to emphasize that the correct black hole thermodynamics can be obtained from a different perspective by analyzing the contribution of the torsional surface term which naturally appears in the TEGR formulation of GR. By ``correct'', we obviously mean that the Schwarzschild black hole entropy ends up being one fourth of its horizon area. Nonetheless, the price to be paid for the correctness of this alternative approach is the choice of reference frames, this is, the specification of the tetrad field, up to LLT solving the equation (\ref{cond}). Then, beyond pointing out the limited range behind the findings in Ref. \cite{Oshita:2017nhn}, the aim of this paper was to properly extend those results by taking into account free variations of the tetrad field on the boundary surface, which are consistent with $\delta g_{\mu\nu}=0$ there. In so doing we made our two main contributions, namely: 

a) The characterization of an infinite number of frames leading to $\mathcal{S}=A/4$ and $\mathcal{E}=M$. These frames belong to the equivalence class of tetrads producing the right variations on the boundary surface, that is to say, leading to $\delta S_{B}=\delta S_{GHY}$. Among them we found radial boosts and some rotations of the static frame, Eqs. (\ref{staticfboo}) and (\ref{staticfrot}) with $\Theta$ given in (\ref{lsefesol}), respectively. Those were our examples, but certainly many more can be found.

b) Because of their adequate asymptotic structure, the frames referred to in (a) ``carry their own background''. Precisely, no background subtraction is needed in order to regularize the divergences occurring as a consequence of a deficient choice of tetrad, as it is the case of the diagonal frame (\ref{diagtet}).

What seems to be relevant in the analysis here performed is the obliged introduction of observers through the presence of a field of tetrads. It is of course very well known that the detection of the black hole temperature and entropy depends on the state of motion of the particle detector \cite{Birrel}, \cite{Ruth}. In this sense, the treatment of black hole thermodynamics by means of the TEGR seems to enrich the usual view because, at least when manifolds with boundary are considered, the torsional boundary term explicitly brings the tetrad into scene by means of the requirement $\delta S_{B}=\delta S_{GHY}$. In the usual GR metric approach, the value $\mathcal{S}=A/4$ corresponding to the entropy perceived by a static observer outside the Schwarzschild black hole horizon, is obtained purely through the Euclidean action constructed from the GHY boundary term, which does not involve the tetrad at all, but only the metric. In the present context, we found a class of observers perceiving the right spacetime energy and entropy, and perhaps on the same footing of relevance, a relatively systematic way of characterize them by means of Eq. (\ref{cond}). Once classified, it would be important to inquire if these ``proper thermodynamic frames'' play any role in the construction of a well-posed action principle in tetrad-GR gravity, and presumably, in its subsequent quantum extension.

\bigskip

{\bf{\emph{Acknowledgments.}}}
This work is supported by ANID Chile through FONDECYT Grant Nº 1220871  (P.A.G., and Y. V.). P.A.G. and Y.V would like to thank Instituto Balseiro (Argentina) and P.A.G to Facultad de Ciencias, Universidad de La Serena (Chile) for their hospitality. F.F. is most thankful to ANID (Chile) for its generous and kind invitation founded through FONDECYT, and to Universidad de La Serena, where most of the research leading to this article took place. He is member of \emph{Carrera del Investigador Cient\'{i}fico} (CONICET, Argentina), and his work is supported by CONICET and Instituto Balseiro (UNCUYO).


\begin{thebibliography}{99}


\bibitem{York:1972sj}
J.~W.~York, Jr.,
\emph{Role of conformal three geometry in the dynamics of gravitation},
Phys. Rev. Lett. \textbf{28} (1972) 1082.


\bibitem{Gibbons:1976ue}
G.~W.~Gibbons and S.~W.~Hawking,
\emph{Action Integrals and Partition Functions in Quantum Gravity},
Phys. Rev. D \textbf{15} (1977) 2752.




\bibitem{Oshita:2017nhn}
N.~Oshita and Y.~P.~Wu,
 \emph{Role of spacetime boundaries in Einstein's other gravity},
Phys. Rev. D \textbf{96} (2017) 044042.
[arXiv:1705.10436 [gr-qc]].

\bibitem{TEGRbook} R. Aldrovandi and J. G. Pereira, \emph{Teleparallel Gravity -- An Introduction}, Springer (2013).

\bibitem{Manuel} M. Hohmann, {\it Teleparallel Gravity}, contribution to the book ``\emph{Signatures and experimental searches for modified and quantum gravity}'' [arXiv:2207.06438  (gr-qc)] (2022).


\bibitem{Martin}
M.~Kr\v{s}\v{s}\'ak,
\emph{Holographic Renormalization in Teleparallel Gravity},
Eur. Phys. J. C \textbf{77} (2017) 44.
[arXiv:1510.06676 [gr-qc]].


\bibitem{Martin2} M. Krššák, \emph{Bulk Action Growth for Holographic Complexity}, [arXiv:2308.04354 (hep-th)] (2023).



\bibitem{Sebastian}
S.~Bahamonde, L.~Ducobu and C.~Pfeifer,
\emph{Scalarized black holes in teleparallel gravity},
JCAP \textbf{04} (2022) 018.
[arXiv:2201.11445 [gr-qc]].


\bibitem{Erdmenger:2022nhz}
J.~Erdmenger, B.~He\ss{}, I.~Matthaiakakis and R.~Meyer,
\emph{Universal Gibbons-Hawking-York term for theories with curvature, torsion and non-metricity},
SciPost Phys. \textbf{14} (2023) 099.
[arXiv:2211.02064 [hep-th]].




\bibitem{Erdmenger:2023}
J.~Erdmenger, B.~He\ss{}, I.~Matthaiakakis and R.~Meyer,
\emph{Gibbons-Hawking-York boundary terms and the generalized geometrical trinity of gravity}, [arXiv:2304.06752 (hep-th)] (2023).



\bibitem{Jose}
J.~Beltr\'an Jim\'enez, L.~Heisenberg and T.~S.~Koivisto,
\emph{Teleparallel Palatini theories},
JCAP \textbf{08} (2018) 039.
[arXiv:1803.10185 [gr-qc]].



\bibitem{Christian1}
C.~G.~Boehmer and E.~Jensko,
\emph{Modified gravity: A unified approach},
Phys. Rev. D \textbf{104} (2021) 024010.
[arXiv:2103.15906 [gr-qc]].




\bibitem{Christian2} C. G. Boehmer and E. Jensko, \emph{Modified gravity: a unified approach to metric-affine models}, J. Math. Phys. \textbf{64} (2023) 082505. [arXiv:2301.11051 (gr-qc)].


\bibitem{Rengo2}
S.~W.~Hawking,
\emph{Quantum Gravity and Path Integrals},
Phys. Rev. D \textbf{18} (1978) 1747.


\bibitem{Rengo3} S.~W.~Hawking, \emph{The path integral approach to quantum gravity}, in General Relativity: an Einstein centenary survey, S.~W.~Hawking and W. Israel eds. (1979) 746. Cambridge University Press.




\bibitem{Maluf1}
J.~W.~Maluf,
\emph{The Gravitational energy-momentum tensor and the gravitational pressure},
Annalen Phys. \textbf{14} (2005) 723.
[arXiv:gr-qc/0504077 [gr-qc]].


\bibitem{Maluf2}
J.~W.~Maluf, F.~F.~Faria and S.~C.~Ulhoa,
\emph{On reference frames in spacetime and gravitational energy in freely falling frames},
Class. Quant. Grav. \textbf{24} (2007) 2743.
[arXiv:0704.0986 [gr-qc]].




\bibitem{FYF}
R.~Ferraro and F.~Fiorini,
\emph{Spherically symmetric static spacetimes in vacuum f(T) gravity},
Phys. Rev. D \textbf{84} (2011) 083518.
[arXiv:1109.4209 [gr-qc]].



\bibitem{Bras}
V.~C.~de Andrade, L.~C.~T.~Guillen and J.~G.~Pereira,
\emph{Gravitational energy momentum density in teleparallel gravity},
Phys. Rev. Lett. \textbf{84} (2000) 4533.
[arXiv:gr-qc/0003100 [gr-qc]].



\bibitem{grupo}
R.~Ferraro and F.~Fiorini,
\emph{Remnant group of local Lorentz transformations in $f(T)$ theories},
Phys. Rev. D \textbf{91} (2015) 064019.
[arXiv:1412.3424 [gr-qc]].
























\bibitem{Rubilar}
Y.~N.~Obukhov and G.~F.~Rubilar,
\emph{Covariance properties and regularization of conserved currents in tetrad gravity},
Phys. Rev. D \textbf{73} (2006) 124017.
[arXiv:gr-qc/0605045 [gr-qc]].
\bibitem{Birrel} N. D. Birrel and P. C. W. Davies, \emph{Quantum fields in curved space}, Cambridge University Press (1982). 

\bibitem{Ruth}
M. Appels, R. Gregory and D. Kubiznak, \emph{Thermodynamics of Accelerating Black Holes}, Phys. Rev. Lett. \textbf{117} (2016) 131303. [arXiv:1604.08812 [hep-th]].









\end{thebibliography}
\end{document}